\documentstyle[preprint,prb,aps]{revtex}
\begin{document}
\draft
\title{Quantum energy flow in mesoscopic dielectric structures}
\author{M. P. Blencowe\cite{miles}}
\address{The Blackett Laboratory, Imperial College of Science, Technology and 
Medicine,
London~SW7~2BZ, United Kingdom}
\date{\today}
\maketitle
\begin{abstract}

We investigate the phononic energy transport 
properties of mesoscopic, suspended dielectric wires.  The Landauer formula for
the thermal conductance is derived and its universal aspects discussed. We 
then determine the  variance of the  energy current in the presence
of a steady state current flow. In the final part, 
some initial results are presented
concerning the nature of the temperature fluctuations of a mesoscopic
electron gas
thermometer due to the absorption and emission of wire phonons.    
   
\end{abstract}


\section{Introduction}

Mesoscopic physics might be defined as the study of certain 
quantum electronic phenomena, normally belonging to the atomic domain, 
which through the use of special  microfabrication techniques are
realised in structures having dimensions ranging from tens of nanometers  
up to micrometers. One consequence of our improving ability 
to directly  probe quantum phenomena  at these scales is the 
increasing relevance of the more nontrivial aspects of quantum mechanics
for the proper explanation of the phenomena, such as the need to include in
the description the measurement process. With further advances in fabrication 
techniques, this trend will continue and we can look forward 
to mesoscopic structures  which display the counterintuitive aspects of quantum
mechanics becoming commonplace.

It should also be possible to fabricate mesoscopic structures in which the 
{\it lattice} degrees of freedom behave in a manifestly nonclassical way.
Phononic analogues of various mesoscopic electron phenomena are an obvious
possibility to consider. For example, we might ask whether the thermal 
conductance of a dielectric wire with sufficiently small cross section will 
exhibit steps of universal magnitude (i.e., expressed, 
apart from a   numerical factor, solely in terms of
  Boltzmann's and Planck's constants)
 analogous to the electronic conductance 
steps observed in quantum wires.\cite{vanwees,wharam} Phononic analogues of
various quantum optical phenomena can also be considered, such as squeezed
phonon states.\cite{hu,garrett}  Phonons may be particularly suited for the 
study
of time-dependent phenomena in the mesoscopic domain. The weakness of the
phonon-phonon interaction at low temperatures 
 and also the ability to fabricate mesoscopic
structures having only a few defects, may allow for the possibility to
track the evolution of non-equilibrium phonon distributions as they
approach thermal equilibrium distributions. Such an investigation might provide new
insights into the longstanding fundamental problem concerning the recovery
of macroscopic irreversibility from the microscopic reversible laws (for a
discussion of this problem in the context of mesoscopic systems, see  Ref.\
\onlinecite{landauer1}).

Phonon-confining mesoscopic structures are not as straightforward to realise as
electron quantum wells, wires etc. For acoustic phonons, there are no perfect
thermal insulators; although confined modes may exist in a heterostructure
consisting of layers of material with different elastic properties, there will 
always be unconfined bulk modes with the same energies. The only solution 
is to use {\it suspended} structures, i.e., structures which are 
physically separated from the substrate for most of their extent.  
An additional challenge is the problem of
probing the phonon dynamics in the suspended structures. For example, in order
to measure the thermal conductance of a suspended nanowire, a way must be 
found in which to heat one end of the wire while keeping the other end at
a fixed temperature and also to measure the temperature difference between 
the two ends. As can be appreciated,
it is rather more difficult to fabricate     
suspended nanostructures integrated with 
ultrasensitive probes  than it is  to fabricate 
conventional heterostructures. Several groups have been involved in 
related work during the past few years, with pioneering studies carried out 
by the Cambridge group\cite{potts} and by Wybourne and coworkers.\cite{seyler}
The recent successful experiments of Roukes and coworkers\cite{tighe}
demonstrate their mastery of the fabrication techniques and 
have  opened up for exploration the field of mesoscopic phonon physics.

In this paper we investigate several phonon phenomena which can in principle
be observed using devices similar to those considered by
Roukes {\it et al}.\cite{tighe}
In Sec.\ II we calculate the mean of the  energy current flowing in
a suspended dielectric wire connected at each end to equilibrium
phonon reservoirs at different temperatures. The Landauer formula for the
thermal conductance is recovered from the mean energy current expression and
the  conditions on the phonon energy spectrum for the observation
of conductance steps determined. In actual dielectric wires the energy spectrum
fails to satisfy the conditions and thus the steps cannot be resolved. 
The temperature dependence of the conductance is then
solved numerically for the special case of a GaAs wire with uniform rectangular
cross section. The main results of this section have also been obtained by 
Angelescu {\it et al.}\cite{angelescu} and by  Rego and Kirczenow\cite{rego}.

In Sec.\ III we calculate the variance of the  energy current in the presence
of a steady state current flow. When the temperatures of the two reservoirs 
coincide, so that the average current flow is zero,
we recover the Johnson-Nyquist noise formula for the 
phonon energy current. 

Practically no mention is made in Secs.\ II and\ III about the ways in which 
the conductance or variance of the current  might actually be measured. 
This is partly remedied in Sec.\ IV, where we consider a model thermometer 
consisting of an electron gas confined to a thin cross sectional slab of the 
wire. We investigate the  temperature fluctuations 
occuring  in this electron gas  
caused by the absorption and emission of phonons.   
The remarkable possibility of detecting {\it single} phonons through
the  temperature fluctuations is a consequence of the very small
volume and hence heat capacity of the electron gas. From the magnitude of a
given temperature fluctuation the energy of the absorbed or emitted phonon 
is known and, thus, there is the possibility for high resolution
phonon spectroscopy. In particular, the energy dependence of 
the phonon transmission probability for a suspended wire can be determined.
We develop some of the necessary theory for describing the statistics of the 
fluctuations and, on the basis of the derived expressions, make some initial 
observations concerning the extraction of the transmission probability energy 
dependence from the fluctuation statistics.

In our calculations we use the
second quantization method. This formalism arises quite naturally when 
quantizing the 
lattice degrees of freedom  and also enables a systematic 
derivation of the thermal conductance, current noise and 
temperature fluctuation
formulae. Although we
do not do so in the present work, it is important to try to rederive these
formulae (particularly the current noise and temperature fluctuation 
formulae) 
using a different approach in 
which the 
phonons are described by propagating, spatially localized 
wavepackets.\cite{landauer2,martin} Such an approach might provide a clearer
picture of what the phonons are `actually doing' in the mesoscopic wires.  
   
\section{The Thermal Conductance}

The model wire structure which we shall consider is shown in Fig.\ \ref{fig1}.
Two very long,  perfect leads 
(i.e., crystalline and with uniform cross section) join a 
central
segment in which the phonon scattering occurs. The scattering may be caused
by any combination of the following: 
a changing cross section, surface roughness, or various internal defects. 
The only restriction we place on the scattering is that it be elastic. 
Phonon-phonon interactions are also neglected. The other ends of the two 
leads are connected to  reservoirs  where the phonon distributions 
are  Bose-Einstein distributions. No scattering occurs at the 
reservoir-lead connections.

Our point of departure is the classical equations of motion for the wire
lattice dynamics and the expression for the classical energy current flowing
in the wire. At Kelvin or lower reservoir temperatures, phonon wavelengths
typically exceed several hundred Angstroms and thus the continuum approximation
can be used for the equations of motion:
\begin{equation}
\rho\partial^2 _t u_i -c_{ijkl}\partial_j \partial_k u_l = 0,
\label{motioneqn}
\end{equation}
where $u_i$ denotes the $i$th component of the displacement field, $\rho$ is
the mass density and 
$c_{ijkl}$ is the elastic modulus tensor. 
The displacement field satisfies the following boundary condition at the wire
surface:
\begin{equation}
\left. c_{ijkl}n_j\partial_k u_l \right|_S  =  0, 
\label{boundarycond}
\end{equation}
where $n_j$ is the $j$th component of the unit vector normal to the 
wire surface $S$. In terms of the displacement field and elastic modulus tensor,
the energy current at a given location $x$ is 
(the $x$-coordinate runs along the length of  the wire): 
\begin{equation}
I(x,t)=-c_{xjkl}\int_A dydz \partial_t u_j \partial_k u_l,
\label{classicalenergy}
\end{equation}
where the integral is over the cross sectional surface $A$ at $x$.

In order to quantize the equations of motion (\ref{motioneqn}), we require
a complete set of normal mode solutions. For a perfect, infinitely long wire,
these solutions can be written in the following form:
\begin{equation}
u_{n, q, i}({\bf r},t)=\frac{1}{\sqrt{2\pi}}e^{-i(\omega_{n, q}t -qx)}
\chi_{n, q,i}(y,z),
\label{perfectmode}
\end{equation}
where $q$ is the longitudinal wavevector along the wire axis and $n$ is the
subband label. It follows from the equations of motion that these solutions
can be chosen to satisfy the orthonormality condition
\begin{equation}
\int d{\bf r}\ u^*_{n, q,i} u_{n', q',i} =\delta_{n  n'} \delta(q-q').
\label{perfectorthonormal}
\end{equation}
In the presence of scattering, we can still  construct  solutions in the leads
using the perfect wire solutions  (\ref{perfectmode}) as follows:
\begin{equation}
 u^1_{n, q,i}=\left\{\begin{array}{ll}
 u_{n, q,i}+ \sum_{n'} u_{n', -q', i}\  t^{11}_{n'  n}(\omega)  & \mbox{lead 1}
\nonumber\\
\sum_{n'}u_{n', q', i}\  t^{21}_{n'  n}(\omega)  & \mbox{lead 2}
\end{array}
\right.
\label{wiremode1}
\end{equation}
and
\begin{equation}
 u^2_{n, q,i}=\left\{\begin{array}{ll}
\sum_{n'}u_{n', -q', i}\  t^{12}_{n'  n}(\omega)  & \mbox{lead 1}
\nonumber\\
 u_{n, -q,i}+ \sum_{n'} u_{n', q', i}\  t^{22}_{n'  n}(\omega)  & \mbox{lead 2,}
\end{array}
\right.
\label{wiremode2}
\end{equation}
where $q, q'>0$. The solutions $u^1_{n, q,i}$ describe waves propagating from 
lead 1
to lead 2, while solutions $u^2_{n, q,i}$ propagate from lead 2 to lead 1.
The absolute value of the scattering matrix element
$t^{b a}_{n' n} (\omega)$ gives 
the fraction of the incident
wave in lead $a$, with frequency $\omega$ and subband label $n$, which is
transmitted/reflected into lead $b$ and subband $n'$. In the sum over $n'$,
the frequency $\omega$ is kept fixed, while  $q'$ is treated as a function of
$n'$ and $\omega$ through the   
condition $\omega_{n',q'}=\omega_{n,q}=\omega$.

From energy conservation, the time average of the energy current $I(x,t)$
should be independent of the position $x$. Substituting into the definition
for the energy current (\ref{classicalenergy}) an arbitrary linear combination
of solutions (\ref{wiremode1}) and (\ref{wiremode2}) and demanding that the
time averaged energy currents in leads 1 and  2 be the same, we obtain the 
following conditions on the scattering matrix elements:
\begin{equation}
\sum_{n''} v_{n'',q''}\ t^{11}_{n'' n}(\omega) t^{11 *}_{n'' n'}(\omega)+
\sum_{n''} v_{n'',q''}\ t^{21}_{n'' n}(\omega) t^{21 *}_{n'' n'}(\omega)=
v_{n,q}  \delta_{n  n'},
\label{condition1}
\end{equation}  
\begin{equation}
\sum_{n''} v_{n'',q''}\ t^{22}_{n'' n}(\omega) t^{22 *}_{n'' n'}(\omega)+
\sum_{n''} v_{n'',q''}\ t^{12}_{n'' n}(\omega) t^{12 *}_{n'' n'}(\omega)=
v_{n,q}  \delta_{n  n'}
\label{condition2}
\end{equation}  
and
\begin{equation}
\sum_{n''} v_{n'',q''}\ t^{11}_{n'' n}(\omega) t^{12 *}_{n'' n'}(\omega)+
\sum_{n''} v_{n'',q''}\ t^{21}_{n'' n}(\omega) t^{22 *}_{n'' n'}(\omega)=
0,
\label{condition3}
\end{equation}  
where $v_{n,q}=\partial\omega_{n,q}/\partial q$ is the group velocity. In the
derivation of these conditions, we require the following very useful  relation
between the group velocity and displacement field:
\begin{equation}
i\pi c_{xijl}\int_A dydz\left(u_{n,q,i} \partial_j u_{n',q',l}^* -
u_{n',q',i}^* \partial_j u_{n,q,l}\right)=\rho \omega_{n,q} v_{n,q}
 \delta_{n n'},
\label{relation}
\end{equation}
where $\omega_{n',q'}=\omega_{n,q}$. This relation is obtained
from the equations of motion (\ref{motioneqn}). Finally,
using Eqs. (\ref{perfectorthonormal}) and (\ref{condition1})-(\ref{condition3}),
it is straightforward to show that the wire mode solutions satisfy the
following orthonormality conditions:
\begin{equation}
\int d{\bf r}\ u^{\sigma *}_{n, q,i}u^{\sigma'}_{n', q',i} 
=\delta_{\sigma \sigma'}\delta_{n  n'} \delta(q-q');\ \ \sigma, \sigma'=1,2.
\label{wireorthonormal}
\end{equation}

We are now ready to quantize. 
In the wire leads, the displacement field operator has the solution
\begin{equation}
\hat{u}_i ({\bf r},t)=\sum_{n,\sigma}\int_0^{\infty} dq 
\sqrt{\frac{\hbar}{2\rho\omega_{n,q}}}\left[\hat{a}_{n,q}^{\sigma} 
u^{\sigma}_{n,q,i}({\bf r},t) +\hat{a}_{n,q}^{\sigma\dag}
u^{\sigma *}_{n,q,i}({\bf r},t)\right],
\label{fieldoperator}
\end{equation}
where the phonon creation and annihilation
operators satisfy the commutation relations
\begin{equation}
[\hat{a}_{n,q}^{\sigma},\hat{a}_{n',q'}^{\sigma' \dag}]=
\delta_{\sigma \sigma'}\delta_{n  n'} \delta(q-q').
\label{accr}
\end{equation}
Substituting the field operator solution (\ref{fieldoperator}) into
the energy current operator
\begin{equation}
\hat{I}=-\frac{1}{2} c_{xjkl}\int_A dydz \left(\partial_t \hat{u}_j 
\partial_k \hat{u}_l + \partial_k \hat{u}_l \partial_t \hat{u}_j\right)
\label{energyoperator}
\end{equation}
and then taking the expectation value of $\hat{I}$ at any location $x$
in leads 1 or 2, we obtain
\begin{equation}
\langle\hat{I}\rangle=\frac{1}{2\pi}\sum_{n,n'}\int_{\omega_{n,0}}^{\infty}
d\omega\ \hbar\omega\ v^{-1}_{n,q} v_{n',q'} t^{21}_{n' n}(\omega)
 t^{21 *}_{n' n}(\omega) \left[n_1 (\omega) -n_2 (\omega)\right],
\label{almostfinalcurrent}
\end{equation}       
where
\begin{equation}
n_{\sigma}(\omega)=\frac{1}{e^{\hbar\omega/k_{B}T_{\sigma}}-1},
\label{boseein}
\end{equation}
with $T_{\sigma}$ the temperature of the reservoir at the end of lead $\sigma$. 
In the derivation of Eq. (\ref{almostfinalcurrent}),  
use is made of  relation (\ref{relation}) and conditions 
(\ref{condition1})-(\ref{condition3}). We also use the following  
creation/annihilation operator expectation values:
\begin{equation}
\langle\hat{a}^{\sigma \dag}_{n,q} \hat{a}^{\sigma'}_{n',q'}\rangle=
n_{\sigma}(\omega_{n,q})\delta_{\sigma \sigma'}\delta_{n n'}\delta(q-q').
\label{ca1expectation}
\end{equation}
Defining
\begin{equation}
T^{21}_{n' n}(E)=v^{-1}_{n,q} v_{n',q'} t^{21}_{n' n}(\omega)
 t^{21 *}_{n' n}(\omega),
\label{transprobdef}
\end{equation}
where $E=\hbar\omega$, we can rewrite (\ref{almostfinalcurrent}) as follows:
\begin{equation}
\langle\hat{I}\rangle=\frac{1}{2\pi\hbar}\sum_{n,n'}
\int_{E_{n,0}}^{\infty}
dE\ E\ T^{21}_{n' n}(E) \left[n_1 (E) -n_2 (E)\right].
\label{finalcurrent}
\end{equation}
This is our key expression for the mean energy current. From  the form of
this expression and
condition (\ref{condition1}), we see that the matrix
$T^{21}_{n' n}(E)$ is naturally interpreted as the probability for a phonon
with energy $E$ in subband $n$ of lead 1 to be transmitted into  subband
$n'$ of lead 2. Eq. (\ref{finalcurrent}) is the starting point for the 
investigations in Refs.\ \onlinecite{angelescu} and \onlinecite{rego}.

When the temperature difference between the reservoirs is small, i.e., 
$|T_1 -T_2|\ll T_1, T_2$, we can expand Eq. (\ref{finalcurrent}) to obtain
the wire thermal conductance:
\begin{equation}
\kappa=\frac{\langle\hat{I}\rangle}{|T_1 -T_2|}=
\frac{\pi k^2_{B} T}{6\hbar}\sum_{n, n'}
\int_{\frac{E_{n,0}}{k_{B}T}}^{\infty}
d\epsilon\ g(\epsilon) T^{21}_{n' n}(\epsilon k_{B} T),
\label{thermalconductance}
\end{equation}    
where $T$ is the average temperature and 
\begin{equation}
g(\epsilon)=\frac{3\epsilon e^{\epsilon}}{\pi^2 (e^{\epsilon}-1)^2}.
\label{gfuncdef}
\end{equation}
Eq. (\ref{thermalconductance}) relates the thermal conductance to
the single phonon transmission probability and thus we call this the 
Landauer expression for the phonon thermal conductance. 

The function 
$g(\epsilon)$ satisfies $\int_0^{\infty}d\epsilon\ g(\epsilon) =1$. 
Therefore,
in the absence of scattering a given subband $n$ contributes
to the 
reduced conductance $\kappa/T$ the {\it universal} quantum 
$\pi k^2_{B}/6\hbar
\approx 9.465\times 10^{-13}\ {\rm W K}^{-2}$
in the limit $E_{n, 0}/k_{B} T \rightarrow 0$.\cite{rego}
Whether or not steps can be resolved in the temperature dependence of
the reduced conductance
 depends on the separation of the subband edges $E_{n+1,0}-E_{n,0}$
and also on the size of the temperature interval over which the integral       
$\int_{E_{n, 0}/k_{B} T}^{\infty}d\epsilon\ g(\epsilon)$ goes from being
much less than one to close to one. A rough criterion can be arrived at by
requiring that the temperature at which the $n$th subband contributes $90\%$ of
a universal quantum be less than the temperature at which the $n+1$th
subband contributes $10\%$ of a universal quantum. This yields the following
condition on the subband edge separation:
\begin{equation}
E_{n+1,0}>14 E_{n,0}.
\label{stepcondition}
\end{equation}
Therefore, in order to resolve the steps, the subband separation would have to
increase by an order of magnitude from one subband to the next! In an actual
wire, the separation  typically goes like $E_{n+1,0}/E_{n,0} \sim (n+1)/n$
and thus the steps cannot be resolved. The same conclusion is reached in
Ref.\ \onlinecite{angelescu} where the possibility of using nonequilibrium,
narrow band phonon distributions to observe the steps is also considered.

In Fig.\ \ref{fig2}, we show the temperature dependence of the reduced thermal
conductance for perfect GaAs wires with uniform, rectangular cross sections of
various dimensions comparable to those used in the experiment of 
Ref.\ \onlinecite{tighe}. The only 
GaAs wire characteristics which are needed in order to determine the 
conductance are the zone-center frequencies $\omega_{n,0}$. 
These can be calculated  using the elegant numerical method developed in 
Ref.\ \onlinecite{nishiguchi}. As expected, there are no steplike features.
There is, however, a plateau for $T\rightarrow 0$ where only phonons in the 
lowest
subband with $E_{n,0}=0$ contribute (see also Ref.\ \onlinecite{rego}).
The  plateau has the value
four in universal  quantum units, a consequence of their being four
basic mode types: dilatational, torsional and two types of flexural 
mode.\cite{nishiguchi}

Of course, in actual wires phonon scattering will occur. For example, the 
reservoirs can be much larger than the wires, with
a sharp decrease in cross section where they join. 
Reservoir phonons approaching the wire with transverse
wavelength component exceeding the cross sectional dimensions of the wire will
be backscattered with high probability and
the resulting suppression in the dielectric wire thermal
conductance at low temperatures may conceal the  plateau described above.
An initial investigation of the consequences for the thermal conductance
of a nonuniform cross section can be found in Refs.\ 
\onlinecite{angelescu} and \onlinecite{rego}.  
In classical wave optics and acoustics,
the same strong reflection
phenomenon occurs for waves traveling
in narrowing waveguides and is called `diffractional
blocking'. This phenomenon is also somewhat analogous to the
situation in an electronic quantum wire when the Fermi level lies below the 
lowest subband edge, so that  electrons can only tunnel 
from one contact region to the other, resulting in an exponential suppression in
the conductance. 

Some closely related work to that described in this section
is in the area of dielectric point contact 
spectroscopy.\cite{bogachek,stefanyi,shkorbatov} 
In fact, diffractional blocking
has already been observed in thermal conductance measurements of 
 point contacts;\cite{shkorbatov} by measuring the temperature at which the
thermal conductance dropped sharply, 
it was possible to estimate the contact diameters which were found to be in the
region of tens of nanometers.  

\section{Energy current noise}

Using the methods developed in the preceding section, it is possible to
calculate more nontrivial quantities characterizing the energy flow
in the wire, such as
the variance of the energy current $(\Delta I)^2=\langle\hat{I}^2\rangle
-\langle\hat{I}\rangle^2$. If we  take the expectation value of 
$[\hat{I}(x,t)]^2$,  we obtain a meaningless divergent result, however.
Given that   we cannot measure the current at 
a precise instant, a more realistic quantity to consider is 
the measured current $I_m$  defined as (see, e.g.,
Ref.\ \onlinecite{yurke}),
\begin{equation}  
I_m (x,t)=\int_{-\infty}^{\infty}d\tau\ H(t-\tau) I(x,\tau),
\label{measuredcurrent}
\end{equation}
where $H(t)$ is a causal filter function satisfying $H(t)=0$ for $t<0$ 
and $\int_{-\infty}^{\infty} dt H(t)=1$.
The expectation value of $[\hat{I}_m(x,t)]^2$ is now finite and well-defined. 
The variance of the measured current is calculated using a similar procedure
to that outlined in the previous section for the mean.
Omitting the details and going directly to the final result, we find
\begin{equation}
(\Delta I_m)^2 \approx \frac{B}{2\pi\hbar}\sum_{n, n'}\int_{E_{n,0}}^{\infty}
dE\ E^2\left\{{\cal T}^{21}_{n' n} \left[ n_1-n_2\right]^2
+T^{21}_{n' n}\left[n_1 (n_2 +1) +n_2 (n_1 +1)\right]\right\},
\label{shotnoise}
\end{equation}
where the transmission probability $T^{21}_{n' n}$ is defined in 
(\ref{transprobdef}) and the transmission matrix ${\cal T}^{21}_{n' n}$ is
defined as follows:
\begin{equation}
{\cal T}^{21}_{n' n} =\sum_{m, m'} \left(v_n v_{n'}\right)^{-1} v_m v_{m'}
t^{21}_{m' n'} t^{21 *}_{m' n} t^{21}_{m n} t^{21 *}_{m n'}.
\label{secondtransdef}
\end{equation}
The constant $B$ is the filter bandwidth:
\begin{equation}
B=\frac{1}{2\pi}\int_{-\infty}^{\infty}d\omega \tilde{H}^2 (\omega),
\label{bandwidth}
\end{equation}
where $\tilde{H}(\omega)=\int_{-\infty}^{\infty}dt e^{-i\omega t} H(t)$.
Approximation (\ref{shotnoise}) is a good one provided that the energy scale
$E=\hbar\omega$ over which $\tilde{H}(\omega)$ is nonzero is small as compared
with the energy scales over which the transmission matrices and 
phonon distributions vary (see, e.g., Ref.\ \onlinecite{yurke}).   

Formula (\ref{shotnoise}) resembles the
electron current variance  formula.\cite{lesovik,buttiker,martin}
(The correspondence is even more direct if the electron 
{\it energy} current
variance  is used for comparison rather than the more commonly
considered charge current variance.) Just
as for the electron case, we see that phonon current noise in the presence of
a nonzero steady state  current (i.e., $T_1 \neq T_2$)  contains more
information concerning the transmission characteristics of the wire
than the thermal conductance.

For the special case where the reservoir temperatures are the same,
the ${\cal T}$ matrix term drops out  and Eq. (\ref{shotnoise}) can be
written as follows:
\begin{equation}
(\Delta I_m)^2 \approx 2B k_{B}T^2 \kappa,
\label{nyquist} 
\end{equation}
where $\kappa$ is the thermal conductance (\ref{thermalconductance}). Thus,
 the equilibrium phonon noise gives the
same information concerning the wire transmission characteristics as the 
thermal conductance. 
We call Eq. (\ref{nyquist}) the 
Johnson-Nyquist noise formula for the phonon energy current. Again,
this formula bears a close resemblance to the electron current Johnson-Nyquist
noise formula.\cite{landauer2,buttiker} When the temperature difference is
nonzero
but small, we see from the form of the phonon
distribution terms in Eq.\ (\ref{shotnoise}) that corrections to the
Johnson-Nyquist equilibrium noise are of second order 
in the temperature difference.

\section{A Mesoscopic Thermometer}

In order to probe the phonon  dynamics of a
wire, some kind of  measuring apparatus is obviously required.
It is important to understand the behaviour of that 
part of the apparatus which interacts directly with the wire phonons, 
so that we can know  just what properties of the phonon system are in 
fact being measured.

As our model measuring system, we consider an electron gas confined
to a thin cross sectional slab of one of the wire leads. The gas density and 
slab thickness are small enough so that the wire phonon current is hardly 
affected by its presence. In other words, a  phonon traversing the gas 
layer has only a very small probability to be absorbed. 
The gas density should also be large enough so that the timescale  for the
electron gas to reach internal thermal equilibrium due to electron-electron 
scattering is much less than the timescale separating consecutive
phonon absorption or emission events. This latter assumption allows us to 
assign a temperature to the electron gas which fluctuates in time due to the 
absorption and emission of
phonons. The remaining part of the apparatus, which we do not describe, 
measures the electron gas temperature with negligible disturbance to the gas.
This measuring system is in fact a closely related idealization of that employed
by  Roukes and coworkers in their experiments.\cite{tighe}

Measuring the wire thermal conductance presents no problem. 
A known constant power
source is supplied to reservoir 1, say, 
while reservoir 2 acts as heat sink  with  known temperature. The
electron gas thermometer is  located at the reservoir end of lead 1 and 
its average temperature measured. The conductance is then just the  
power divided by the difference between the gas thermometer temperature and 
the temperature of reservoir 2 [see Eq. (\ref{thermalconductance})].

The  fluctuations of the electron gas temperature
 give much more
information  concerning the wire phonon dynamics than  the 
average temperature. 
The  real possibility to detect temperature fluctuations is a consequence,
as the following estimates show, of the very small electron gas
volume which can be achieved.
For a nearly degenerate electron gas the specific heat is  
approximately
\begin{equation} 
\frac{\partial E}{\partial T}\approx\frac{\pi^2 n V k^2_B T }{2 E_F},
\label{specificheat}
\end{equation}
where $n$ and $V$ are the electron gas number density and volume, respectively.
Using the relation between $E_F$ and $n$ for free electrons
to eliminate $E_F$, Eq. (\ref{specificheat}) becomes
\begin{equation}
\frac{\partial E}{\partial T}\approx
\frac{\pi^{2/3}m n^{1/3} V k^2_B T}{3^{2/3}\hbar^2}.
\label{specificheat2}
\end{equation}
If the electron gas absorbs or emits a thermal phonon with energy $3 k_B T$,
then from (\ref{specificheat2}) we get an approximate temperature change
\begin{equation}
\delta T\approx \frac{3^{5/3}\hbar^2}{\pi^{2/3} m n^{1/3} V k_B}.
\label{tempchange}
\end{equation}
For GaAs with $n=10^{18}\ {\rm cm}^{-3}$, this gives
\begin{equation}
\delta T\approx 0.4 V^{-1}\ {\rm mK},
\label{gaastempchange}
\end{equation}
where the volume $V$ is given in units $\mu{\rm m}^3$. Thus, for an 
an electron gas thermometer with submicron dimensions (which can be
achieved with present fabrication techniques\cite{tighe}), 
absorption or emission
of a thermal phonon will produce a temperature fluctuation in excess of a 
milliKelvin.

Note, however, that it is not possible to measure the 
energy current fluctuations using the electron gas thermometer. 
Although the
energy of an absorbed phonon can be determined from the size of the 
temperature fluctuation, all information is lost concerning the direction
in which the wire phonon was travelling. To gain an initial idea about 
what information  can be obtained concerning the phonon dynamics, we shall
now present some of the theory for the temperature fluctuations.

Much of the theory of photoelectric light detection in quantum optics 
(see, e.g., Chapter 14 of Ref.\ \onlinecite{mandel}) can be
adapted to our present problem. As a phonon detector, however, 
the electron gas thermometer behaves in a more nontrivial manner than the 
photoelectric detector. Unlike a conventional photoelectric detector, 
the gas thermometer can not only detect phonons, but  
measure their energy as well. Furthermore, the gas thermometer can 
emit phonons. When successive phonon detections are correlated, 
these  properties can  make the calculation of various detection 
probabilities rather difficult. In the following, we shall neglect the
correlations. This then allows us to recover all statistical 
properties of the temperature fluctuations from the detection probability 
for very short time intervals  (i.e., short enough 
so that the probability is much less than one).
It should be born in mind, however, that many of the expected interesting
quantum properties will be correlation effects and, thus, it is important
to try to include the correlations in future improvements of the theory. 

The quantity of interest, then, is the probability $R(E,E')\delta t$
that the electron gas,
initially with total energy $E$, has energy $E'\neq E$ after a short time 
interval $\delta t$, due to the absorption or emission of a phonon with energy
$\Delta=|E' -E|$. Recall that we are  assuming the electron gas to
be in internal thermal equilibrium between absorption/emission events.
For a large number of electrons, the electron gas temperature $T$ can be 
determined to good approximation 
from the total energy $E$ of the electron gas by using the relation 
$E=2\sum_{\alpha}\varepsilon_{\alpha} f(\varepsilon_{\alpha})$, where 
$\varepsilon_{\alpha}$ is a single electron energy eigenvalue
and $f(\varepsilon_{\alpha})$ is the Fermi-Dirac distribution.
Neglecting
correlations, an energy probability distribution $P(E)$ 
 will evolve in time according to the following 
equation:
\begin{equation}  
\frac{\partial P}{\partial t}(E,t)=\sum_{E'} P(E',t) R(E',E)
-\sum_{E'} P(E,t) R(E,E').
\label{rateequation}
\end{equation}  
Thus, knowing the rate $R(E,E')$ allows us to in principle determine how a 
probability distribution evolves.  
 
Using  the methods of, e.g., Chapter 14 of Ref.\ \onlinecite{mandel},  we obtain
the following expression for the rate:
\begin{eqnarray}
R(E,E\pm\Delta) & = &\frac{\pi}{\rho\Delta}
\sum_{{\alpha,\beta\atop\varepsilon_{\beta}-
\varepsilon_{\alpha}=\pm\Delta}}
f_{\alpha}(1-f_{\beta})\nonumber\\
& \times &\left[\sum_{n,n'} v^{-1}_{n,q}
\left( |\lambda^{\alpha\beta}_{n,q}|^2 +
|\lambda^{\alpha\beta}_{n,-q}|^2 +\lambda^{\alpha\beta *}_{n,q}
\lambda^{\alpha\beta}_{n',-q'} t^{11}_{n' n} +
\lambda^{\alpha\beta}_{n,q}
\lambda^{\alpha\beta *}_{n',-q'} t^{11 *}_{n' n}\right) 
\left(n_1 +{\case1/2}\mp{\case1/2}\right)\right.\nonumber\\
& + &\left.\sum_{n,n',n''}v^{-1}_{n,q}\lambda^{\alpha\beta}_{n',-q'}
\lambda^{\alpha\beta *}_{n'',-q''} t^{12}_{n' n} t^{12 *}_{n'' n}
(n_2 -n_1)\right],
\label{microrate}
\end{eqnarray}
where all the phonon quantities are evaluated at $\omega=\Delta/\hbar$. This
rate expression is for the case where the thermometer is located in lead 1. 
To obtain the corresponding expression when it is located in lead 2, the lead
indices `1' and `2' should be interchanged wherever they appear. 
The quantity $\lambda^{\alpha\beta}_{n,q}$ is the
electron-phonon matrix element:
\begin{equation}
\lambda^{\alpha\beta}_{n,q}=\int_V d{\bf r}\psi^*_{\beta}({\bf r})\left(\Xi_d
\partial_i u_{n,q,i}({\bf r}) +\frac{\overline{e}e_{14}}{4\pi\epsilon}
\int_V d{\bf R}\frac{e^{-q_0 |{\bf r}-{\bf R}|}}{|{\bf r}-{\bf R}|}
\partial_{(12}^2u_{n,q,3)}({\bf r})\right)\psi_{\alpha}({\bf r}),
\label{ephmatrixelement}
\end{equation}       
where the integrals are over the electron gas volume, $\psi_{\alpha}$ is the
electron energy eigenstate and $U_{n,q,i}$ is the phonon mode solution 
in the lead
[see Eq. (\ref{perfectmode})]. The first term in the large brackets
is the deformation component of the potential
and the second term is the piezoelectric
component (see, e.g., Chapter 3 of Ref.\ \onlinecite{ridley}).

The rate expression (\ref{microrate}) comprises two terms, of which the second
involving the $t^{12}$  matrix is the most interesting. A
possible procedure would be to measure
the rates for $T_1 \neq T_2$ and also for $T_1 =T_2$, with $T_1$ the same in 
each 
case. The difference between the two rates would then be given by just the 
second
term.
Because this term is proportional to  $t^{12} t^{12 *}$ evaluated at 
$\omega=\Delta/\hbar$, the rate difference therefore provides direct information
concerning the energy dependence of the phonon scattering matrix averaged
over the various subbands.
Of course, knowledge of the electron-phonon matrix elements would be required
in order to extract this information. 

\section{Conclusion}

We have presented several results concerning the energy flow properties of
mesoscopic, suspended dielectric wires. The  mean of
the energy current was calculated and a Landauer formula for the thermal
conductance obtained. When scattering is absent, 
each phonon subband contributes a universal quantum $\pi k^2_{B}/6\hbar$
to the reduced conductance $\kappa/T$. Steps are not observed, however,
because of the broadness of the Bose-Einstein distribution as compared with
the subband edge separation. The temperature dependence 
of the reduced conductance was solved numerically for the example of a
GaAs wire with uniform, rectangular cross section. The  variance of the
energy current was then calculated and the Johnson-Nyquist equilibrium 
noise formula obtained as a special case. In the final part, an initial
investigation was carried out concerning the nature of
the  fluctuations of a mesoscopic electron gas thermometer due to
the absorption and emission of wire phonons. It was found that the fluctuations
give direct information concerning the energy dependence of the phonon 
scattering matrix for the wire.

\acknowledgments

The author would like to thank A. MacKinnon, N. Nishiguchi,
 T. Paszkiewicz,  M. L. Roukes, 
and T. N. Todorov for helpful and stimulating discussions.
He would also like to thank the organizers of the 
`Probing Nanoelectronic Structures Using Phonons EC Network Meeting', Bangor, 
Sept. 1997, for providing the opportunity to present some of this work.
Funding was provided by the EPSRC under Grant No. GR/K/55493.

\begin{figure}
\caption{Schematic diagram of the model wire. 
The left and right  reservoirs are at temperatures $T_1$ and $T_2$, 
respectively.}
\label{fig1}
\end{figure}

\begin{figure}
\caption{Reduced thermal conductance versus temperature for perfect GaAs wires 
with uniform  rectangular cross section $200\ {\rm nm}\times 400\ {\rm nm}$
(solid line), $200\ {\rm nm}\times 300\ {\rm nm}$ (dashed line)
and $200\ {\rm nm}\times 100\ {\rm nm}$ (dotted line). 
The reduced conductance is given
in units $\pi k^2_{B}/6\hbar
\approx 9.465\times 10^{-13}\ {\rm W K}^{-2}$.}
\label{fig2}
\end{figure}

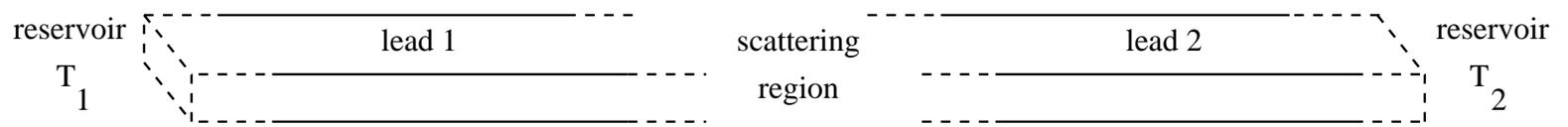
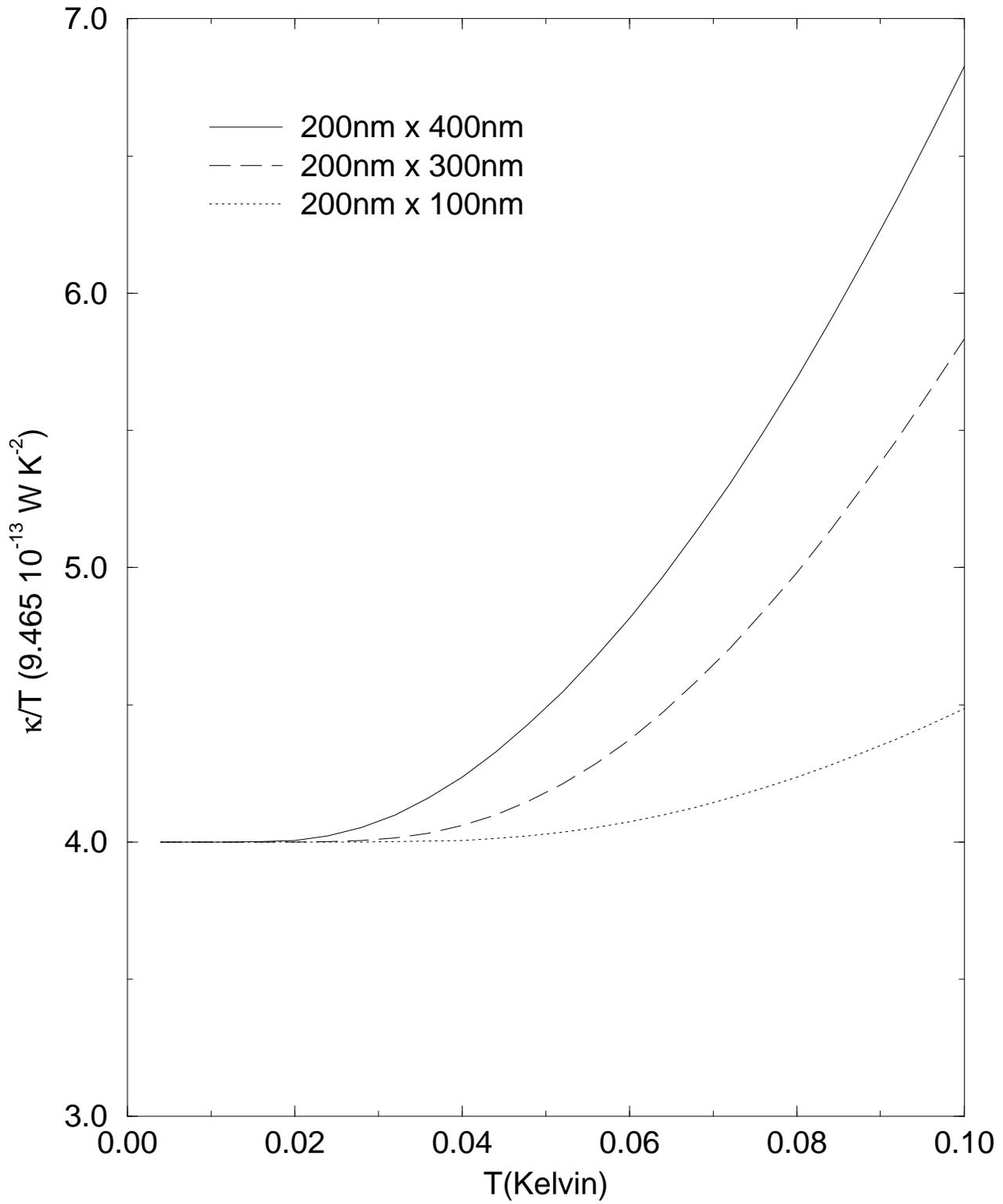
\end{document}